\documentclass[aps,prl,reprint,showpacs,superscriptaddress]{revtex4-1}
\usepackage{amssymb,amsmath,amsbsy,graphicx,hyperref,xcolor}

\begin{document}
\title{Spin-Valley Filtering in Strained Graphene Structures with Artificially Induced Carrier Mass and Spin-Orbit Coupling}

\author{Marko M. Gruji\'c}\email{marko.grujic@etf.bg.ac.rs}
\affiliation{School of Electrical Engineering, University of Belgrade, P.O. Box
3554, 11120 Belgrade, Serbia} \affiliation{Department of Physics, University of
Antwerp, Groenenborgerlaan 171, B-2020 Antwerp, Belgium}
\author{Milan \v{Z}. Tadi\'c}\email{milan.tadic@etf.bg.ac.rs}
\affiliation{School of Electrical Engineering, University of Belgrade, P.O. Box
3554, 11120 Belgrade, Serbia}
\author{Fran\c{c}ois M. Peeters}\email{francois.peeters@uantwerpen.be}
\affiliation{Department of Physics, University of Antwerp, Groenenborgerlaan
171, B-2020 Antwerp, Belgium}

\begin{abstract}
The interplay of massive electrons with spin-orbit coupling in bulk graphene
results in a spin-valley dependent gap. Thus, a barrier with such properties
can act as a filter, transmitting only opposite spins from opposite valleys. In
this Letter we show that a strain induced pseudomagnetic field in such a barrier
will enforce opposite cyclotron trajectories for the filtered valleys, leading
to their spatial separation. Since spin is coupled to the valley in the
filtered states, this also leads to spin separation, demonstrating a
spin-valley filtering effect. The filtering behavior is found to be
controllable by electrical gating as well as by strain.
\end{abstract}
\maketitle

Graphene is considered a promising material for future spintronic applications,
in part due to its long spin relaxation length \cite{pesin12,tomb07,zomer12}.
Furthermore, owing to its band structure with two inequivalent valleys, $K$ and
$K^{\prime}$, it has revived the field of valleytronics \cite{rycerz07,neto09}.
The low energy excitations in the two valleys behave as Dirac-Weyl particles,
which is most famously manifested in the presence of a magnetic field, in which
Landau levels scale as $\sqrt{B}$, with a unique level at zero energy
\cite{neto09,grujic11}. Besides, it is known that straining graphene causes
time-reversal invariant gauge fields to appear, i.e., an effective magnetic
field with opposite signs in opposite valleys, providing a tool for
manipulating the valley degree of freedom \cite{neto09}. Recent experiments
demonstrated large values of this pseudomagnetic field, which could hardly be
matched in practical applications by real magnetic fields \cite{levy10}.

In this Letter we study the transmission through a thin 1D graphene barrier
with artificially induced mass and spin-orbit coupling (SOC), in the presence
of a pseudomagnetic field using the continuum approach. Our motivation for
studying such a structure is twofold. In part it is due to a shift to a new
paradigm in 2D materials research, whereby their properties are custom tailored
according to specific needs by stacking different 2D crystals on top of each
other. These are the so-called van der Waals heterostructures \cite{geim13}.
More importantly, and in the light of this paradigm, recent theoretical and
experimental work suggests that mass and SOC, which are vanishing in intrinsic
graphene, could be induced with appropriate substrates and/or adatom deposition
\cite{zhong11,sachs11,kinder12,amet13,hunt13,neto2009,weeks11,alicea12,ozy13,jin13,ozy14}.
The studied device is found to behave as a spin-valley filter, thus lying in
the intersection of the fields of spintronics and valleytronics.

In the continuum approach the carrier mass is captured by a staggered potential
term $\Delta$, while SOC is captured by a masslike term $\Delta_{SO}$. The
presence of both will result in a competition to open topologically distinct
gaps \cite{kane05}. This competition reflects on the gap size given by
$2\left|s\tau\Delta_{SO}+\Delta\right|$, where $s=+1/-1$ labels the spin
$\uparrow/\downarrow$ and $\tau=+1/-1$ labels the valley $K/K^{\prime}$ degrees
of freedom \cite{ezawa13}. Thus, for different spin-valleys different gaps can
arise. In order to get some insight into the problem, we first study
transmission through a barrier with a real magnetic field. Regardless of the
magnetic field, whenever $\Delta_{SO}\neq0$ and $\Delta\neq0$, there is an
energy range where $s\tau=+1$ states are suppressed, while $s\tau=-1$ states
are not. In other words only one spin from one valley, and the opposite spin
from the opposite valley are transmitted. The main effect of the magnetic field
is to impose restrictions on incident angles over which the transmission can
occur. This is caused by the cyclotron orbits, which are the same for all spins
and valleys.

We subsequently apply the pseudomagnetic field, which leads to the reversal of
the effective field, and the effective cyclotron orbits in one of the valleys.
This provides the benefit of spatially separating the transmitted states
according to their valley degree of freedom, and accordingly their spin degree
of freedom as well. Thus a combined spin-valley filter can be obtained.
Furthermore, we show that chemical potential and strain can act as a switch,
rendering control over the filtering behavior. Filtering behavior in graphene
devices was studied before
\cite{chaves10,masir11,wu11,dean12,jiang13,myoung13,tsai13,lu13,yoko13,yoko14}; however the
mechanism proposed in this Letter is novel, and previously unexplored.
Practical implications are discussed at the end of the Letter.

Our starting point is the Dirac-Weyl equation, in the presence of mass, SOC,
and a magnetic field perpendicular to the sheet, $B_z$. In this case we choose
the Landau gauge $\mathbf{A}=\left(0,A_y\right)$, and the Dirac-Weyl
Hamiltonian reads

\begin{equation}\label{dirac-weyl}
H=\hbar v_F\left[\tau k_x\sigma_x+(k_y+\frac{e}{\hbar}A_y)\sigma_y\right]+s\tau\Delta_{SO}\sigma_z+\Delta\sigma_z,
\end{equation}
where $v_F$ is the Fermi velocity, and $\sigma_z$ is a Pauli matrix operating
in the sublattice subspace.

We use the parameter $\tau_B$, such that $B_z=\tau_B B$, to capture the
valley-dependent nature of the pseudomagnetic field. Setting $\tau_B=+1$ models
the influence of the real magnetic field, while $\tau_B=\pm\tau$ models the two
types of the pseudomagnetic field. The (pseudo)magnetic field, mass, and SOC
exist only in the barrier of width $W$. The vector potential is
therefore given by

\begin{equation}
A_y=\begin{cases}0, &x<0\\\tau_BBx, &0\leq x\leq W\\\tau_BBW, &x>W\end{cases}.
\end{equation}

In the chosen Landau gauge $k_y$ is a good quantum number and the solutions
have the form $\Psi(x,y)=\exp(ik_yy)\left(\psi_A(x),\psi_B(x)\right)^T$.
Introducing $\hbar v_F\epsilon=E$, and $\hbar
v_F\delta=s\tau\Delta_{SO}+\Delta$, and decoupling the system, in the barrier
one obtains

\begin{equation}\label{difjna}
\left[\partial_x^2\mp\tau\tau_B\frac{1}{l_B^2}-(k_y+\tau_B\frac{x}{l_B^2})^2+\epsilon^2-\delta^2\right]\psi_{A/B}=0,
\end{equation}
where $l_B=\sqrt{\hbar/e B}$. Using the transformation
$z=\sqrt{2}\left(k_yl_B+\tau_Bx/l_B\right)$, the solutions are expressed in
terms of the parabolic cylinder functions $D_{\nu}(z)$ (see the Supplemental
Material \cite{supmat} for details), and read

\begin{equation}\label{resenje}
\psi_{II}=C_1\left(\begin{array}{c}D_{\nu_A}(z)\\gD_{\nu_B}(z)\end{array}\right)+C_2\left(\begin{array}{c}D_{\nu_A}(-z)\\-gD_{\nu_B}(-z)\end{array}\right),
\end{equation}
where $\nu_{A/B}=\left(\epsilon^2-\delta^2\right)l_B^2/2\mp\tau\tau_B/2-1/2$,
and

\begin{equation}\label{g}
g=i\left[\frac{\sqrt{2}}{\left(\epsilon+\tau\tau_B\delta\right)l_B}\right]^{\tau\tau_B}.
\end{equation}

On the other hand, the incident wave function is

\begin{equation}
\psi_I=e^{ik_xx}\left(\begin{array}{c}1\\
\tau e^{i\tau\phi}\end{array}\right)+re^{-ik_xx}\left(\begin{array}{c}1\\ \tau e^{i\tau(\pi-\phi)}\end{array}\right),
\end{equation}
while the solution in the third region reads

\begin{equation}
\psi_{III}=t\sqrt{\frac{k_x}{k_x^{\prime}}}e^{ik_x^{\prime}x}\left(\begin{array}{c}1\\ \tau e^{i\tau\theta}\end{array}\right).
\end{equation}

Here, $\phi=\arctan k_y/k_x$ and $\theta=\arctan k_y^{\prime}/k_x^{\prime}$
denote the energy propagation directions before and after the barrier, where
$k_y=q_y\left(0\right)$ and $k_y^{\prime}=q_y\left(W\right)$, while
$q_y\left(x\right)=\epsilon\sin\phi+eA_y\left(x\right)/\hbar$ is the effective
transverse momentum. The longitudinal momenta before and after the barrier are
given by $k_x=\epsilon\cos\phi$ and $k_x^{\prime}=\epsilon\cos\theta$
\cite{supmat}. Note that all these expressions are valid for the valence band
as well \cite{supmat}. Matching the wave functions at the interfaces gives a
system of equations, whose solution yields the transmission amplitude $t$

\begin{equation}\label{tampl}
t=\frac{2g\tau\cos\phi\left(G_A^+G_B^-+G_A^-G_B^+\right)}{e^{ik_x^{\prime}W}f}\sqrt{\frac{k_x^{\prime}}{k_x}},
\end{equation}
where

\begin{equation}
\begin{split}
f=g^2\left(F_B^+G_B^--F_B^-G_B^+\right)+e^{i\tau\left(\theta-\phi\right)}\left(F_A^+G_A^--F_A^-G_A^+\right)\\
+g\tau e^{i\tau\theta}\left(F_B^-G_A^++F_B^+G_A^-\right)+g\tau e^{-i\tau\phi}\left(F_A^+G_B^-+F_A^-G_B^+\right).
\end{split}
\end{equation}
Here the coefficients $F^{\pm}$ and $G^{\pm}$ are given by

\begin{align}
F^{\pm}_{A/B}&=D_{\nu_{A/B}}\left[\pm\sqrt{2}k_yl_B\right],\\
G^{\pm}_{A/B}&=D_{\nu_{A/B}}\left[\pm\sqrt{2}\left(k_yl_B+\tau_B\frac{W}{l_B}\right)\right].
\end{align}

\begin{figure}
\centering
\includegraphics[width=8.6cm]{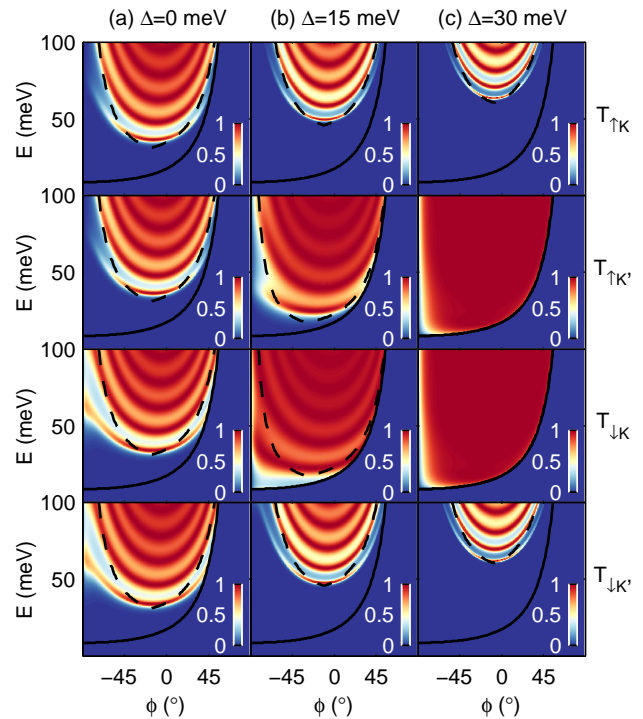}
\caption{Contour plots of transmission coefficient versus energy and incident
angle, for all spin and valley flavors. $\Delta_{SO}$ equals $30$ meV, while
$\Delta$ is varied: (a) $\Delta=0$, (b) $\Delta=15$ meV and (c) $\Delta=30$ meV.
The width of the barrier is taken to be $W=100$ nm, and $B=0.2$ T.}
\label{fig1}
\end{figure}

In Fig. \ref{fig1} we look at the behavior of transmission coefficients
($T=\left|t\right|^2$) in detail for a real magnetic field ($\tau_B=+1$). Here
we show contour plots of, from top to bottom, $T_{\uparrow K}$, $T_{\uparrow
K^{\prime}}$, $T_{\downarrow K}$ and $T_{\downarrow K^{\prime}}$, as a function
of incident energy and angle. We adopt a set of parameters that illustrates our
main points clearly: $\Delta_{SO}=30$ meV, $W=100$ nm and $B=0.2$ T, whereas
$\Delta$ varies from $0$ in (a), to $\Delta=15$ meV in (b) and $\Delta=30$ meV
in (c).

A common feature of all the cases depicted in Fig. \ref{fig1} is that
transmission is forbidden outside the transmission window delineated by the
solid black line. This is because the magnetic field enforces cyclotron motion,
resulting in asymmetric transmission curves with respect to the incidence angle
\cite{martino07,masir08}. This boundary is obtained by requiring that the
longitudinal momentum after the barrier becomes imaginary, so that only
evanescent waves can exit, and therefore no transmission can occur. The
longitudinal momentum in the third region is given by
$k_x^{\prime2}=\epsilon^2-q_y\left(W\right)^2$. Hence, this window is
determined by a critical energy, below (above) which the transmission is not
possible

\begin{equation}\label{prozor}
\epsilon_{cr1}^{c/v}=\frac{\pm\gamma}{1\mp\tau_B\sin\phi},
\end{equation}
where $\gamma=W/l_B^2$, and $c$ ($v$) denotes the conduction (valence) band. The
window depends on $W,B$ and $\phi$, i.e., it is not a function of
$\Delta_{SO},\Delta,s$ or $\tau$ at all, as can be observed in Fig. \ref{fig1}.
However, the transmission within this window obviously depends on
$\Delta_{SO},\Delta,s$ and $\tau$.

As already mentioned, in the presence of mass and SOC, the bulk band gap is
given by $2\left|s\tau\Delta_{SO}+\Delta\right|$. Therefore, when both
parameters are present, the $s\tau=+1$ states experience a larger gap than the
$s\tau=-1$ states. To see how this might reflect on the transmission through a
barrier we need to examine the behavior of the quasiclassical momentum within
the barrier $q_x\left(x\right)=\sqrt{\epsilon^2-\delta^2-q_y\left(x\right)^2}$
\cite{supmat}. Therefore, through the appearance of $\delta$, the
quasiclassical momentum depends on $\Delta_{SO},\Delta,s$ and $\tau$. More
specifically, when both $\Delta$ and $\Delta_{SO}$ are nonzero, whether
classically forbidden regions inside the barrier will appear depends crucially
on the product $s\tau$, which is a clear manifestation of the bulk band gap.
The existence of forbidden regions in the barrier does not necessarily imply
that the momentum after the barrier is imaginary. To see this, one can express
the critical energy below (above) which the former happens

\begin{equation}\label{bar}
\epsilon_{cr2}^{c/v}=\pm\max\left(\frac{\pm\tau_B\gamma\sin\phi+\sqrt{\gamma^2+\delta^2\cos^2\phi}}{\cos^2\phi},\frac{\left|\delta\right|}{\cos\phi}\right).
\end{equation}

This critical boundary is drawn in dashed black lines in Fig. \ref{fig1}, and it
coincides with the transmission window (Eq. \eqref{prozor}) only when the bulk
band gap is closed. Therefore, in between $\epsilon_{cr1}^c$ and
$\epsilon_{cr2}^c$ transmission is possible, but only by tunneling through the
forbidden region (regions) in the barrier, and thus perfect transmission cannot
occur. Above the $\epsilon_{cr2}^c$ boundary, however, there is no attenuation
within the barrier, and the resulting transmission is determined by the
interference of electron waves. It is important to point out that below the
minimum of $\epsilon_{cr2}^c$, which coincides with the bottom of the
conduction band, the transmission is strongly suppressed.

One issue requires clarification. For the case $\Delta=0$, shown in Fig. \ref{fig1}(a),
$\epsilon_{cr2}^c$ is the same for all spin and valley flavors. However, the
transmissions for spin up and spin down are obviously different. This
discrepancy arises due to the factor $g$, appearing in the transmission
amplitude, Eq. \eqref{tampl}. This factor is in turn just a reflection of the
form of the Landau level (LL) eigenstates \cite{supmat}. In fact one can easily
show that the solution given by Eq. \eqref{resenje} reduces to the LL
eigenstates once the incident energy is equal to a particular LL \cite{supmat}.

It is known that inversion symmetry breaking can lead to the appearance of
magnetic moments coupled with the valley degree of freedom, which in turn
influence the LLs \cite{xiao07}. Similar moments arise when SOC is present as
well, albeit coupled with the spin degree of freedom \cite{supmat}. It is these
moments that cause spin-distinguished transmission found in Fig. \ref{fig1}(a).
A similar behavior occurs when only $\Delta$ is nonzero, but with valley
differentiation instead. In fact, we have found that all of the contour plots
obey the symmetry $\Delta_{SO}\leftrightarrow\Delta$, $s\leftrightarrow\tau$.
This stems from the fact that the band gap and the magnetic moments display the
same symmetry as well \cite{supmat}. We stress, however, that this behavior has
little to no impact on the effect we describe here, and will be studied in
detail elsewhere.

Introducing $\Delta$ will cause shrinking (enlarging) of the evanescent region
for $s\tau=-1$ ($s\tau=+1$), Fig. \ref{fig1}(b). This will lead to the appearance of an energy
range where only $s\tau=-1$ states are not suppressed. Furthermore note that
these states also display lower fringe constrast.
This is because the barrier is effectively reduced for these
states. Finally, for the case $\Delta_{SO}=\Delta$, depicted in column (c),
$s\tau=+1$ states are even further suppressed. On the other hand, for
$s\tau=-1$ the barrier vanishes, as the effective
dispersion returns to a Dirac cone. These states are influenced only by the magnetic field \cite{martino07,masir08},
which can also be inferred from the fact that now
$\epsilon_{cr1}^{c/v}=\epsilon_{cr2}^{c/v}$. This means that they experience no
reflection at the walls of the barrier and as a consequence there are no
resonances.

Therefore, as long as $\Delta_{SO}\neq0$ and $\Delta\neq0$, in a particular
energy range only spin up states from the $K$ valley and spin down states from
the $K^{\prime}$ valley are transmitted. Introducing a pseudomagnetic field, by
for instance setting $\tau_B=+\tau$, means that the effective magnetic field in
$K^{\prime}$ valley flips. This in turn flips the transmission window in this
valley to $\epsilon_{cr1K^{\prime}}^c=\gamma/\left(1+\sin\phi\right)$. Thus,
spatial separation of the states from each valley will occur, which is an
obvious consequence of their opposite cyclotron trajectories. Furthermore,
since spin is coupled to the valley degree of freedom in the transmitted
states, this will inevitably lead to spin separation as well.

\begin{figure}
\centering
\includegraphics[width=8.6cm]{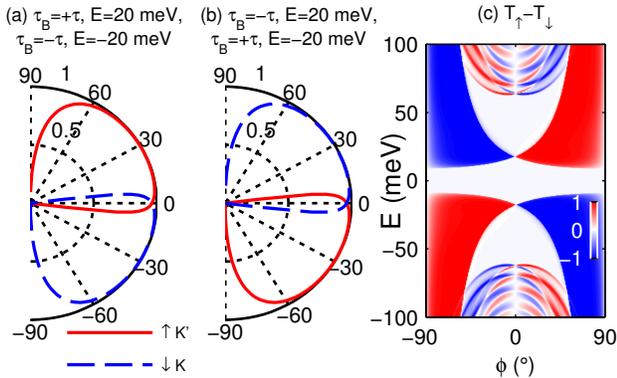}
\caption{Polar plots of the transmission coefficient versus the incident angle for various
strains and energies. In (a) $\tau_B=\pm\tau$, $E=\pm20$ meV give the same
transmission, while in (b) $\tau_B=\mp\tau$, $E=\pm20$ meV give the same
transmission. All other parameters are the same as in Fig. \ref{fig1}(c). The
electric control of the spin-valley filtering is clearly seen in (c), where the
contour plot of $T_{\uparrow}-T_{\downarrow}$ is shown.}
\label{fig2}
\end{figure}

Additionally, it follows from Eq. \eqref{tampl} that the transmission
coefficient for $-\phi$, $-s$, $-\tau$ equals the one for $\phi$, $s$, $\tau$,
which is a manifestation of time-reversal symmetry \cite{footnote1}. In other
words, the transmission for spins in the valley where the effective magnetic
field is reversed, will just be a mirror image of the transmission from the
opposite valley and opposite spin, for which the effective magnetic field
stayed the same. This is displayed in Fig. \ref{fig2}(a), for the same set of
parameters as in Fig. \ref{fig1}(c), and $E=20$ meV, where the spin-valley
filtering behavior is apparent. On the other hand, by choosing the opposite
strain, $\tau_B=-\tau$, the effective magnetic field will be flipped in both
valleys. This will lead to flipping of the filtered spin and valley, as
depicted in Fig. \ref{fig2}(b), since both transmission windows flip (see Eq.
\eqref{prozor}). In other words strain could act as a switch \cite{footnote2}.

Furthermore, the switching can also be achieved by controlling the chemical
potential instead of strain. To see this, note that the transmission window for
a given spin and valley in the valence band $\epsilon_{cr1}^v$ is a mirror
reverse of the one in the conduction band $\epsilon_{cr1}^c$, Eq.
\eqref{prozor}. This is a consequence of different cyclotron trajectories for
electrons and holes, and the same symmetry is obeyed by the semiclassical
critical boundary, given in Eq. \eqref{bar}. Moreover, since
$T\left(\epsilon,\tau_B=+\tau\right)=T\left(-\epsilon,\tau_B=-\tau\right)$
holds, Figs. \ref{fig2}(a) and (b) also correspond to $\tau_B=-\tau$, $E=-20$
meV and $\tau_B=\tau$, $E=-20$ meV, respectively \cite{supmat}. The effect of
controlling the chemical potential on spin filtering is depicted in Fig.
\ref{fig2}(c), where the outlines of the transmission windows can be clearly
seen. Note that the same plot holds for $T_{K^{\prime}}-T_K$, albeit with
opposite filtering in the overlap region of both transmission windows.
Therefore, the control of the transmitted spin-valley outside of the
transmissionless gap $\left[-\gamma/2,\gamma/2\right]$ could be established by
means of electrical gating. Additionally, there exist optimal energy ranges for
filtering in the valence and conduction band, $\left[-\gamma,-\gamma/2\right]$
and $\left[\gamma/2,\gamma\right]$, respectively, where the transmitted states
do not overlap \cite{supmat}.

Finally, we include some practical considerations. First note that only minor
straining would be required for inducing a pseudomagnetic field of $0.2$ T in a $100$ nm wide barrier, given the strain pattern described in Ref. \cite{guinea10}.
Since $\Delta_{SO}$ and $\Delta$ equal zero in graphene, these two parameters
would have to be induced artificially in the barrier, a feat possible because
bulk electrons are fully exposed on the graphene surface. Hexagonal boron
nitride (hBN) has an intrinsically broken inversion symmetry, and forms a
generally higher quality electronic heterostructure with graphene as opposed to
other substrates \cite{dean10}, manifested in reduced charge impurities,
ultraflatness, and high electron mobility. It also has a minuscule lattice
mismatch with respect to graphene \cite{zhong11}, which causes a moir\'{e}
pattern, resulting in a Hofstadter fractal spectrum \cite{ponomarenko13,dean13}.
While the emerging superlattice potential was suggested to induce insulating
puddles with opposing masses \cite{sachs11,zarenia12}, it was also argued that
an average gap should be opened nevertheless \cite{kinder12}. Recently, a gap
of about $30$ meV in a graphene/hBN composite, consistent with inversion
symmetry breaking was detected \cite{hunt13,woods14}. The average gap appears
because the area of the favored commensurate stacking expands by stretching of
the graphene lattice, once the two layers are well aligned
\cite{amet13,woods14,jung14}.

On the other hand, it was suggested that engineering SOC in graphene can be
achieved by adatoms or substrates \cite{neto2009,weeks11,alicea12,jin13}. This
was indeed experimentally verified recently, where SOC as high as $17$ meV was
observed \cite{ozy13,ozy14}. Since SOC in Eq. \eqref{dirac-weyl} commutes with
out-of-plane spin, increasing it will not affect scattering of this spin
component. However, inversion symmetry breaking will cause new extrinsic spin
relaxation mechanisms \cite{pesin12,ochoa12}. The use of hBN as a substrate
would prove beneficial here, since it was shown that the resulting
heterostructure supports very long spin relaxation lengths \cite{zomer12}.
Moreover, we argue that scattering processes could also be reasonably reduced
by manipulating barrier length and/or strain patterns.

In conclusion, we proposed a device that enables filtering and spatial separation of
opposite spin-valley pairs. The proposed spin-valley filter consists of a
strained barrier with artificially engineered electron mass and SOC. Nanoribbon
geometry could provide the practical testing ground for this effect, with the
barrier formed perpendicular to the ribbon. If $\Delta>\Delta_{SO}$, the device
would be in the topologically trivial phase, and the polarized current could in
principle be detected by leads attached to the edges of the ribbon. On the
other hand, if $\Delta_{SO}>\Delta$, edge states could become a nuisance.
However the device could still operate in the domain of electron optics. In
other words, the effect would be observable for a sufficiently collimated beam
injected far from the edges. Collimation could also be achieved by means of a
smooth Klein barrier in front of the studied device \cite{che06}.

\begin{acknowledgements}This work was supported by the Serbian Ministry of
Education, Science, and Technological Development, the Flemish Science
Foundation (FWO-Vl), and the Methusalem program of the Flemish government.
\end{acknowledgements}


\begin{thebibliography}{00}
\bibitem{footnote1}In \cite{supmat} we demonstrate that
    $T_{\theta}\left(\alpha\right)=T_{\phi}\left(-\alpha\right)$, allowing a straightforward
    interpretation of Fig. \ref{fig2} in terms of the exiting angle as well.
\bibitem{footnote2}Note that the symmetry
    $\tau_B=+\tau,\phi\rightarrow\tau_B=-\tau,-\phi$ does not hold in
    general, due to the forementioned interplay with the orbital magnetic
    moments, although it effectively appears so.
\end{thebibliography}
\end{document}